    \documentclass[12pt]{article}
    \usepackage{psfig}  \usepackage{epsfig}  \usepackage{graphics}
    \usepackage{inputenc}
    \inputencoding{latin1}
     \newlength{\dinwidth}                       
     \newlength{\dinmargin}                      
\newcommand{\ZETF}[3]{{\it Zh.~Eksp.~Teor.~Fiz.} {\bf #1} ({#3}) {#2}}

\newcommand{\YF}[3]{{\it Yad.~Fiz.} {\bf #1} ({#3}) {#2}}
\newcommand{\SJNP}[3]{{\it Sov.~J.~Nucl.~Phys.} {\bf #1} ({#3}) {#2}}
\newcommand{\JETP}[3]{{\it Sov.~Phys.~JETP} {\bf #1} ({#3}) {#2}}
\newcommand{\NPB}[3]{{\it Nucl.~Phys.} {\bf B#1} ({#3}) {#2}}

\newcommand{\ZPC}[3]{{\it Z.~Phys.} {\bf  C#1} ({#3}) {#2}}

\newcommand{\CPC}[3]{{\it Comput.~Phys.~Comm.} {\bf #1} ({#3}) {#2}}

\def\lsim{\mathrel{\rlap{\lower4pt\hbox{\hskip1pt$\sim$}}
    \raise1pt\hbox{$<$}}}                
\def\gsim{\mathrel{\rlap{\lower4pt\hbox{\hskip1pt$\sim$}}
    \raise1pt\hbox{$>$}}}                
\def\ariadne{A\scalebox{0.8}{RIADNE}}
\def\lepto{L\scalebox{0.8}{EPTO}}
\def\herwig{\scalebox{0.8}{HERWIG}}
\begin{document}
\begin{flushright}
  LU TP 99--22\\
  hep-ph/9908368 \\
  August 1999
\end{flushright}
\vspace*{10mm}
\begin{center}  \begin{Large} \begin{bf}
      The Colour-Dipole model and the\\
      \ariadne\ program at high $Q^2$\\
    \end{bf}  \end{Large}
  \vspace*{5mm}
   {\it To be published in the proceedings of the\\
   HERA Monte Carlo workshop, Hamburg, Germany, 1998-1999}

   \vspace*{5mm}
  \begin{large}
    Leif L\"onnblad\\
  \end{large}
  Department of Theoretical Physics, Lund University,\\
  Helgonav\"agen 5 S-223~62 Lund, Sweden
\end{center}
\begin{quotation}
\noindent
{\bf Abstract:} I present a modification of the Colour Dipole Model
for DIS as implemented in \ariadne\ to better describe events at high
$Q^2$.
\end{quotation}

\section{Introduction}

Since the start of the HERA machine, the \ariadne\ \cite{ariadne}
program has generally been regarded the event generator giving the
best overall description of the measured DIS hadronic final states.
For most observables it gives a better reproduction of experimental
results than more conventional generators such as \herwig\ 
\cite{herwig} and \lepto\ \cite{lepto} which implements initial- and
final-state parton showers based on DGLAP evolution \cite{DGLAP}. This
has in particular been true in the region of small $x$ and $Q^2$, and
it has often been taken as an indication that DGLAP evolution is
inappropriate for this region. The fact that \ariadne\ has the feature
of \textit{non-ordering of transverse momenta} in common with the BFKL
evolution \cite{BFKL}, has also been taken as an indication that the
latter is a more correct description of small-$x$ evolution

As the experimental statistics increased also for events at high
$Q^2$, where we assume that our description of the underlying physics
is on more solid theoretical grounds, it became apparent that the
\ariadne\ program was unable to describe this data \cite{hzhiq2}. Even
in the so-called Current Breit Hemisphere (CBH), where the events are
expected to look much like half an event in an e$^+$e$^-$ collision at
the same $Q^2$ \cite{eden}, the \ariadne\ program gave far too little
transverse momentum, despite the fact that the same program gives a
very good description of e$^+$e$^-$ data e.g.\ from LEP.

In this paper I will comment on the reason of this poor description of
data at high $Q^2$, and present a modification of the underlying
Colour-Dipole Model (CDM) for DIS \cite{cdm}. The resulting improvement
in the reproduction of measured data is presented elsewhere in these
proceedings.

\section{The problem}

In the CDM, it is assumed that all gluon radiation in a DIS event can
be described in terms of dipole radiation between the struck quark and
the proton remnant, much in the same way as the radiation in a
hadronic e$^+$e$^-$ collision can be described as dipole radiation
from the initial quark and anti-quark. The main difference is that,
while in e$^+$e$^-$ the initial $q\bar{q}$-pair is essentially
point-like, the proton remnant in a DIS event is an extended object.

\begin{figure}[htb]
  \begin{center}
    \epsfig{figure=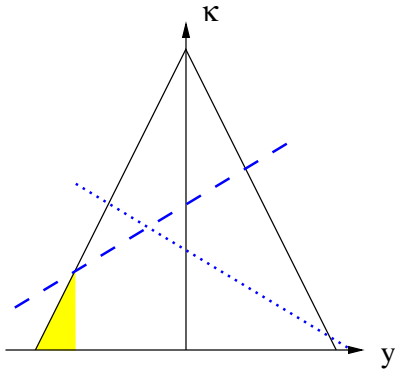}
  \end{center}
\caption[junk]{{\it
    The phase space available for final state particles in DIS in the
    $(\kappa,y)$ plane. $\kappa=\ln(q_\perp^2/\Lambda^2)$, $y$ is
    rapidity in the hadronic CMS. }}
\label{fig:phase1}
\end{figure}

Looking at the phase space available for gluon radiation in a DIS
event (conveniently described as an approximately triangular area in
the plane of rapidity and logarithm of the transverse momentum of the
emitted gluon in figure \ref{fig:phase1}), it is ultimately limited by
the momenta (given here in light-cone components) of the incoming
virtual photon $(Q_+,-xP_-,\vec{0})$ and proton $(0, P_-,\vec{0})$. In
the emission of a gluon with momentum $(q_+,q_-,\vec{q}_\perp)$, the
phase space restriction is given by $q_+=q_\perp e^y < Q_+$ and
$q_-=q_\perp e^{-y} < P_-$. But since the proton remnant is an
extended object, say with transverse extension $1/\mu\sim1$~fm, it is
reasonable to assume that the gluon only can access a fraction of the
negative light-cone momentum of the remnant. Just as radiation of
short wavelengths from an extended antenna is suppressed, we get an
extra phase space restriction from the condition that the gluon with a
transverse wavelength $\propto1/q_\perp$ can only resolve a fraction
$(\mu/q_\perp)^\alpha$ of the remnants momentum (where $\alpha$ is the
dimensionality of the remnant)
\begin{equation}
  q_\perp e^{-y}<(\mu/q_\perp)^\alpha P_-,
  \label{eq:phase1}
\end{equation}
corresponding to the dotted line in figure \ref{fig:phase1}.

For small values of $Q^2$ the struck quark can no longer be considered
point like, and one can argue that emissions of gluons with $q_\perp >
Q$ should also be suppressed, since the gluon only can resolve a
fraction $\propto (Q/q_\perp)^{\alpha'}$ of the positive light-cone
component of the virtual photon. This gives an extra phase-space
restriction corresponding to the dashed line in figure
\ref{fig:phase1}.  In this way the concept of a resolved virtual
photon is present in \ariadne.

\begin{figure}[htb]
  \begin{center}
    \epsfig{figure=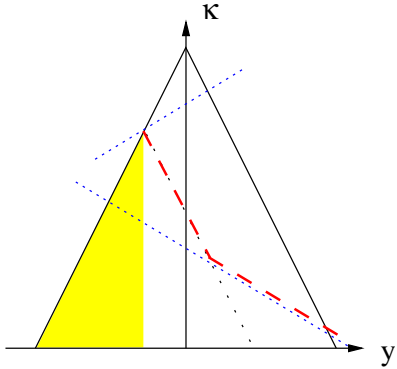}
  \end{center}
\caption[junk]{{\it
    The phase space available for final state particles in DIS at high
    $Q^2$. }}
\label{fig:phase2}
\end{figure}

At high $Q^2$, we expect the radiation in the CBH, which corresponds
to the shaded area in figure \ref{fig:phase2}, to look very much like
half an e$^+$e$^-$ event. But the phase space restriction due to the
extension of the proton remnant actually cuts away part of this
region.  This is most likely the reason why \ariadne\ has given too
little radiation in high-$Q^2$ events.

\section{The solution}

To solve this deficiency in the model we note that in the quark-parton
model we have a collision between a virtual photon and an incoming
quark with momentum $(0,xP_-,\vec{0})$. So in some sense a fraction x
of the incoming protons momentum has already been localized, and this
fraction should be easily available for gluon emission. We can then
rewrite the condition in eq.~(\ref{eq:phase1})
\begin{equation}
  \label{eq:phase2}
  q_\perp e^{-y}< \max(x,(\mu/q_\perp)^\alpha)P_-,
\end{equation}
and the restriction on the phase space would instead correspond to the
dashed line in figure \ref{fig:phase2}.

It could, of course, be argued that the whole of the negative
light-cone momentum $xP_-$ of the incoming quark is needed to put the
quark on-shell, due to the virtuality of the incoming photon with a
negative light-cone momentum of $-xP_-$. And that to radiate an extra
gluon, more negative light-cone momenta has to be accessed from the
proton remnant. But we should remember that the picture of the proton
remnant, where the momentum is spread out evenly over its transverse
extension, certainly is an oversimplification. It is not unnatural to
assume that in the vicinity where the struck quark was localized, the
momentum is more concentrated.  In any case it is clear that the
restriction in eq.~(\ref{eq:phase1}) is too hard, and seems like
eq.~(\ref{fig:phase2}) is a reasonable modification. Indeed, as
reported in \cite{hiq2data}, it seems that this modification is
approximately what is needed to make CDM reproduce measured data at
high $Q^2$.

\section{Conclusions}

The modification of the colour dipole model for DIS presented here
does fix the most serious problems with the reproduction of data at
high $Q^2$ for \ariadne. All problems are, however, not completely
solved. Elsewhere in these proceedings \cite{hiq2data}, it has been
proven to be difficult to find a parameter set which at the same time
can describe general event shapes and jet observables. Jet shapes seem
to be particularly difficult to describe with \ariadne. Whether this
can be solved with small adjustments or if it is due to an inherent
flaw in the model remains to be seen.

\end{document}